\documentclass[preprint,showpacs,showkeys,preprintnumbers,aps]{revtex4}

\usepackage{graphicx}
\usepackage{dcolumn}
\usepackage{bm}
\newcommand{\Exp}[1]{\textrm{e}^{#1}}
\begin{document}

\preprint{KEK-TH-853}

\title{Momentum dependence of the spectral functions in the $\mathcal{O}(4)$ \\linear sigma model at finite temperature}

\author{Yoshimasa HIDAKA}
 \email{hidaka@post.kek.jp}

\author{Osamu MORIMATSU}
 \email{osamu.morimatsu@kek.jp}
\author{Tetsuo NISHIKAWA}
 \email{nishi@post.kek.jp} 
\affiliation{%
Institute of Particle and Nuclear Studies, High Energy Accelerator Research Organization, 1-1, Ooho, Tsukuba, Ibaraki, 305-0801, Japan
}

\date{\today}

\begin{abstract}
The spatial momentum dependence of the spectral function for $\pi$ and 
$\sigma$ at finite temperature is studied by employing the $\mathcal{O}(4)$ linear sigma model and adopting a resummation technique called optimized perturbation theory (OPT).
The poles of the propagators are also searched for.

We analyze the spatial momentum dependence of the imaginary part of the self-energy and find its temperature-dependent part to vanish in the large momentum limit.
This is because the energy of the particles in the heat bath which participate in the process becomes large and therefore the Bose distribution function vanishes.

We then calculate the spectral functions and search for the poles of the propagators.
First, we discuss the temperature dependence for zero spatial momentum.
We reproduce the spectral functions in both $\pi$ and $\sigma$ channels in the previous work and find that the pole, which is responsible for the threshold enhancement of the spectral functions in both $\pi$ and $\sigma$ channels, is different from the one for the peak at zero temperature.
Secondly, we discuss the spatial-momentum dependence.
When the temperature is low the spatial momentum dependence is also small.
As the temperature becomes higher the spatial momentum dependence becomes also large.
When the spatial momentum is smaller than the temperature, the spectral functions do not deviate much from those at zero spatial momentum.
 Once the spatial momentum becomes comparable with the temperature, the deviation becomes considerable.
As the momentum becomes much larger than the temperature, thermal effects effectively decrease.
\end{abstract}
\maketitle

\section{Introduction} 
One of the current interests in the strong interaction physics 
is how the properties of hadrons change 
at finite temperature and/or density.
To study this has become increasingly important
since it is closely related to the data obtained 
in currently carried and future planned 
high energy heavy ion collision experiments \cite{QM:2001}.

A novel feature at finite temperature and/or density is the absence of Lorentz symmetry due to the existence of the heat bath and/or matter and therefore, the spectral functions of hadrons depend on two rotationally invariant variables
in addition to temperature and/or density, for example four momentum squared and spatial momentum squared.
Since the hadron spectra observed in the heavy ion collision experiments 
have various spatial momenta,
it is important to know theoretically how the spectral functions
depend on the spatial momentum in order to draw a conclusion about the change of the properties of hadrons at finite temperature and/or density.

The purpose of the present paper is to study the spatial momentum dependence of the spectral functions and the pole of the propagator in the complex four momentum square plane for hadrons at finite temperature.
We employ a low energy effective theory of QCD, $\mathcal{O}(4)$ linear sigma model and adopt a resummation technique called optimized perturbation theory (OPT) to it \cite{Chiku:1998kd}.
As is well known, naive perturbative expansion breaks down at finite temperature and proper resummation of higher order terms is necessary \cite{Weinberg:1974hy,Dolan:1974qd,Kirzhnits:1976ts}. 
By calculating the self energies of $\pi$ and $\sigma$ in the OPT to one-loop and analyzing their spectral functions, we study how the spectral functions depend on the temperature and the spatial momentum.
We also search for the poles of the propagators.

The properties of $\sigma$ at finite temperature at rest were studied in \cite{Chiku:1998kd,Patkos:2002xb,Patkos:2002vr,Dobado:2002xf}. 
The authors in \cite{Chiku:1998kd,Patkos:2002xb} analyzed the spectral function in $\mathcal{O}(4)$ linear sigma model. 
By adopting OPT Chiku and Hatsuda found threshold enhancement of the spectral function due to the partial restoration of chiral symmetry.
Patkos et al. applied large N expansion, and obtained results similar to Chiku and Hatsuda.
Dobado et al. studied the mass and width using chiral unitary approach.
They searched for the poles of the T-matrix of $\pi\pi$ scattering and showed that the width grows with temperature, while the mass decreases slightly.

This paper is organized as follows.
In the second section, we give some basics of $\mathcal{O}(4)$ linear sigma model and thermal field theory
and briefly review the formalism of OPT.
Then the origin of the momentum dependence is discussed in the third section.
The fourth section is devoted to show the numerical results of the analysis.
Finally we summarize this paper in the fifth section.
The explicit expressions of one-loop self energies at finite temperature 
are given in the appendix.
\section{Formulation} 
    \subsection{Linear Sigma Model} 
    \label{sec:LSM}
    
The Euclidean Lagrangian density of the $\mathcal{O}(4)$ linear sigma model is given by
    \begin{equation}
        \mathcal{ L}_E = \frac{1}{2}[(\partial\vec{\phi})^2+\mu^2\vec{\phi}^2]
        +\frac{\lambda}{4!}(\vec{\phi}^2)^2-h\phi_0+\textrm{(counter term)},
        \label{eq:LSMLagrangian}
    \end{equation}
where $\vec{\phi}(x) = (\phi_0,\vec{\pi})$, and $h\phi_0$ is an explicit chiral symmetry breaking term.
When the $\mathcal{O}(4)$ symmetry is spontaneously broken, $\phi_0$ has an expectation value:
    \begin{equation}
    \langle \phi_0 \rangle_{\beta} = \xi,
    \end{equation}
where $<>_\beta$ is the thermal average, and $\beta=\frac{1}{T}$.
After taking out the expectation value, as $\phi_0= \sigma+\xi$, one can rewrite the Lagrangian density Eq.(\ref{eq:LSMLagrangian}) as
    \begin{eqnarray}
    \mathcal{L}_E 
    &=& \frac{1}{2}(\partial\sigma)^2+\frac{1}{2}m_{0\sigma}^2\sigma^2+\frac{1}{2}(\partial\vec{\pi})^2+\frac{1}{2}m_{0\pi}^2\vec{\pi}^2+\frac{\lambda}{4!}\sigma^4+\frac{\lambda}{4!}(\vec{\pi}^2)^2\nonumber\\
    && +\frac{\lambda}{12}\sigma^2\vec{\pi}^2+\frac{\lambda\xi}{6}\sigma\vec{\pi}^2+\frac{\lambda\xi}{6}\sigma^3+(\frac{\lambda\xi^3}{6}+\mu^2\xi-h)\sigma+\frac{\xi^2}{2}\mu^2+\frac{\lambda\xi^4}{4!}-h\xi\\
    && +\textrm{(counter term)}.\nonumber
    \end{eqnarray}
Here, $m_{0\sigma}$ and $m_{0\pi}$ are the tree-level masses:
    \begin{equation}
    m_{0\sigma}^2=\mu^2+\frac{\lambda}{2}\xi^2 \textrm{,~~~~~~~ } m_{0\pi}^2=\mu^2+\frac{\lambda}{6}\xi^2.
    \end{equation}
$\xi$ is determined by the condition $\langle \sigma \rangle_{\beta}=0$ which reads in the one-loop order as
    \begin{eqnarray}
    &&m_{0\pi}^2\xi-h+\frac{\lambda\xi}{32\pi^2}(m_{0\sigma}^2(\ln\left|\frac{m_{0\sigma}^2}{\kappa^2}\right|-1)+m_{0\pi}^2(\ln\left|\frac{m_{0\pi}^2}{\kappa^2}\right|-1)) \nonumber\\
    & + &  \frac{\lambda\xi}{4\pi^2}\int^{\infty}_0 dkk^2(\frac{n(\omega_{\sigma})}{\omega_{\sigma}}+\frac{n(\omega_{\pi})}{\omega_{\pi}})=0,
    \label{eq:static_condition}
\end{eqnarray}
where $\kappa$ is the renormalization point, $\omega_{\pi}=\sqrt{\bm{k}^2+m_{0\pi}^2}$, $\omega_{\sigma}=\sqrt{\bm{k}^2+m_{0\sigma}^2}$ and $n(\omega)=\frac{1}{\Exp{\beta\omega}-1}$.
\subsection{Thermal Field Theory} 
In the imaginary-time formalism of the thermal field  theory \cite{Matsubara:1955ws}, the Matsubara propagator is defined by
\begin{equation}
 \Delta(i\omega_n,\bm{p},T)=\int_0^\beta d\tau e^{i\omega_n\tau}\int d^3x e^{-i\bm{p}\cdot\bm{x}}<T_\tau\phi(-i\tau,\bm{x})\phi(0,\bm{0})>_\beta,
\end{equation}
where
\begin{equation}
T_\tau\phi(-i\tau,\bm{x})\phi(0,\bm{0})=\theta(\tau)\phi(-i\tau,\bm{x})\phi(0,\bm{0})+\theta(-\tau)\phi(0,\bm{0})\phi(-i\tau,\bm{x}),
\end{equation}
and $\omega_n$ is the Matsubara frequency, $\omega_n=2\pi nT$ (n is an integer).
The Matsubara propagator is defined for discrete pure imaginary energies only but
one can define the propagator, $\Delta(p_0,\bm{p},T)$, for any continuous complex energy $p_0$,  as the proper analytic continuation of the Matsubara propagator \cite{Kirzhnits:1976ts}.
Then, the spectral function at finite temperature, $\rho(p,T)$, is obtained as the discontinuity of the propagator:
\begin{equation}
    i\rho(p,T) = \Delta(p_0+i\eta,\bm{p},T)-\Delta(p_0-i\eta,\bm{p},T).\hspace{1cm} (\eta\rightarrow 0^+),\nonumber\\
\end{equation}
Introducing the self-energy $\Pi(p_0,\bm{p},T)$ by
\begin{equation}    
    \Delta(p_0,\bm{p},T)=\frac{-1}{p^2-m^2-\Pi(p_0,\bm{p},T)},
\label{eq:Matsubaraprop}
\end{equation}
one can write the spectral function as
\begin{equation}
    \rho(p,T) = \frac{-2\textrm{Im}\Pi(p_0+i\eta,\bm{p},T)}{(p^2-m^2-\textrm{Re}\Pi(p_0+i\eta,\bm{p},T))^2+(\textrm{Im}\Pi(p_0+i\eta,\bm{p},T))^2}.
\end{equation}

\subsection{OPT} 
It is known that naive perturbation breaks down at $T \neq 0$, and resummation of higher order terms is necessary \cite{Weinberg:1974hy,Dolan:1974qd,Kirzhnits:1976ts}.
We adopt here a resummation technique called optimized perturbation theory (OPT) \cite{Chiku:1998kd}.
The idea of the OPT consists of dividing the mass parameter of the Lagrangian into two pieces, $\mu^2=m^2-(m^2-\mu^2)\equiv m^2-\chi$, and treating the one, $m^2$, as the tree-level mass term while the other, $\chi$, as perturbation.

The Lagrangian of the linear sigma model Eq.(\ref{eq:LSMLagrangian}) can be rewritten as follows:
\begin{eqnarray} 
\mathcal{ L}_E &=&\frac{1}{2}[(\partial\vec{\phi})^2+m^2\vec{\phi}^2]
    -\frac{1}{2}\chi\phi^2+\frac{\lambda}{4!}(\vec{\phi}^2)^2-h\phi_0+\textrm{(counter term)}.
\end{eqnarray}
The thermal effective potential is written as the functional integral:
\begin{equation}
V(\vec{\varphi},T)=\frac{\ln\int\mathcal{D}\phi\exp[-\frac{1}{\delta}\int^{\beta}_0d^4x(\mathcal{L}(\vec{\phi}+\vec{\varphi})+\vec{J}\cdot\vec{\varphi})]}{\int^{\beta}_0d^4x}.
\end{equation}
Here, $J=\frac{\partial V}{\partial\varphi}$, $\int^{\beta}_0d^4x=\int^{\beta}_0d\tau\int d^3x$, and the parameter, $\delta$, is introduced for convenience $(\delta=1)$.
The OPT is defined by the expansion in $\delta$ with the assignment, $m^2=O(\delta^0)$, $\chi=O(\delta)$ while the naive loop expansion corresponds to the assignment, $m^2=O(\delta^0)$, $\chi=O(\delta^0)$.

When spontaneous symmetry breaking takes place, tree level masses are
    \begin{equation}
m_{0\pi}^2=m^2+\frac{\lambda}{6}\xi^2\textrm{,~~~~~~~~~}m_{0\sigma}^2=m^2+\frac{\lambda}{2}\xi^2.
\end{equation}
The parameter $\xi$ is determined by the stationary condition:
\begin{equation}
    \left.\frac{\partial V(\vec{\varphi},T,m^2)}{\partial\vec{\varphi}}\right|_{\vec{\varphi}=(\xi,\bm{0})}=0.
\end{equation}
Note that the derivative with respect to $\xi$ does not act on $m^2$.
This condition is equivalent to the condition $\langle \sigma\rangle_{\beta}=0$ in Eq.(\ref{eq:static_condition}).
If any Green's function is  calculated in all orders in OPT, they should not depend on the arbitrary mass, $m$. However they depend on the arbitrary mass, $m$, if one truncates perturbation series at certain order.
One can determine this arbitrary parameter so that the correction terms
 are as small as possible.
We adopt the following condition \cite{Chiku:1998kd}:
\begin{equation}
\left[p^2-m^2_{0\pi}-\{{\Pi}_{\pi}^0(p)+{\Pi}_{\pi}^T(0,T)\}\right]_{p=(m_{0\pi},\bm{0})}=0,
\label{eq:FAC}
\end{equation}
where ${\Pi}_{\pi}^0(p)$ and ${\Pi}_{\pi}^T(p,T)$ are, respectively, $T$ independent part and $T$ dependent part of the self-energy for $\pi$. 
Explicit forms of $\sigma$ and $\pi$ self-energy at one-loop level are given in Appendix. 
\subsection{Physical Conditions}
The renormalized parameters $\mu^2$, $\lambda$ and $h$ can be fixed by three physical conditions at zero temperature.
We use the following three conditions \cite{Chiku:1998kd}.\\
(1)The on-shell condition for the pion:
$\Delta_{\pi}^{-1}(p^2=m_{\pi}^2,T=0)= 0$ or $\Pi_{\pi}(p^2=m_{\pi}^2,T=0)=0$, where $m_{\pi}=m_{0\pi}=140$MeV.\\
(2)Partially conserved axial-vector current(PCAC) relation: $f_{\pi}m_{\pi}^2=h\sqrt{Z_{\pi}}$. $Z_{\pi}$ is the finite wave function renormalization constant which depends on the renormalization scheme,
\begin{equation}
Z_{\pi}=\frac{1}{1-\frac{\partial}{\partial p^2}\Pi_{\pi}(p^2=m_{\pi}^2,T=0)}
\end{equation}
 $f_{\pi}$ is the pion decay constant, $f_{\pi}=93$MeV.\\
(3)The peak energy of the spectral function in the sigma channel is determined from the $\pi\pi$ scattering experiment to be $p^2=(550\textrm{MeV})^2$. \\
The parameters given by solving these condition are $\mu^2=-(284\textrm{MeV})^2$, $\lambda=73.0$, $h=(123\textrm{MeV})^3$, $\kappa=255\textrm{MeV}$.

\subsection{Analytic Continuation}
%
From now on we regard the propagator as a function of $p^2$, $|\bm{p}|$ and $T$, i.e. $\Delta(p^2,|\bm{p}|,T)$.
To know the location of the poles of the propagator in the complex $p^2$ plane helps us to understand the behaviour of the spectral function.
For this purpose it is necessary to obtain the analytically continued propagator in the unphysical sheet.
Therefore, from Eq.(\ref{eq:Matsubaraprop}), the problem is reduced to performing analytic continuation of the self-energy into the unphysical sheet.
The self-energy has branch cuts on the real axis, $p^2<(m_{0\sigma}-m_{0\pi})^2$ and $p^2>(m_{0\sigma}+m_{0\pi})^2$ for $\Pi_{\pi}(p^2,|\bm{p}|,T)$ and $p^2<0$ and $p^2>4m_{0\pi}^2$ for $\Pi_{\sigma}(p^2,|\bm{p}|,T)$.
Analytic continuation of the function through different cuts define different analytically continued functions on different Riemann sheets.
Here, we discuss analytic continuation of $\Pi_{\pi}(p^2,|\bm{p}|,T)$ through the cut, $0<p^2<(m_{0\pi}-m_{0\sigma})^2$ as an example.
Our method is the same as the one in ref.\cite{Patkos:2002vr}.
In the physical plane one may define $\Pi_{\pi}(p^2,|\bm{p}|,T)$ by the following $I^{(2)}(p^2,|\bm{p}|,m_{0\pi},m_{0\sigma},T)$ introduced in Appendix,
\begin{eqnarray}
I^{(2)}(p^2,|\bm{p}|,m_{0\pi},m_{0\sigma},T) &=& \frac{1}{16\pi^2}[\frac{1}{\bar{\epsilon}}+2-a_{\pi}\ln(\frac{m_{0\pi}^2}{\kappa^2})-a_{\sigma}\ln(\frac{m_{0\sigma}^2}{\kappa^2}) - c\ln\frac{(c+a_{\pi})(c+a_{\sigma})}{(c-a_{\pi})(c-a_{\sigma})}]\nonumber\\
&& -\int^{\infty}_0\frac{dkk}{16\pi^2|\bm{p}|}
                           \left(\frac{n(\omega_{\pi k})}{\omega_{\pi k}}\ln\frac{((a_{\pi}p^2+k|\bm{p}|)^2-p_0^2\omega_{\pi k}^2)}{((a_{\pi}p^2-k|\bm{p}|)^2-p_0^2\omega_{\pi k}^2)}\right.  \nonumber \\
                       &&   \left. +\frac{n(\omega_{\sigma k})}{\omega_{\sigma k}}\ln\frac{((a_{\sigma}p^2+k|\bm{p}|)^2-p_0^2\omega_{\sigma k}^2)}{((a_{\sigma}p^2-k|\bm{p}|)^2-p_0^2\omega_{\sigma k}^2)}\right),
\end{eqnarray}
where
\begin{eqnarray}
&&a_{\pi}=\frac{1}{2}\left\{1-\frac{m_{0\sigma}^2-m_{0\pi}^2}{p^2}\right\},\hspace{1cm}
a_{\sigma}=\frac{1}{2}\left\{1-\frac{m_{0\pi}^2-m_{0\sigma}^2}{p^2}\right\},\nonumber\\
&&c=\frac{1}{2}\sqrt{\left\{1-\frac{(m_{0\sigma}+m_{0\pi})^2}{p^2}\right\}\left\{1-\frac{(m_{0\sigma}-m_{0\pi})^2}{p^2}\right\}}.
\label{eq:parameter}
\end{eqnarray}

This function is analytic in the complex physical $p^2$ plane except for the real axis.
Let $I_{unphys}^{(2)}(p^2,|\bm{p}|,m_{0\pi},m_{0\sigma},T)$ be the analytic continuation of $I^{(2)}(p^2,|\bm{p}|,m_{0\pi},m_{0\sigma},T)$ into the unphysical sheet through the cut $0<p^2<(m_{0\pi}-m_{0\sigma})^2$ and $F(p^2,|\bm{p}|,m_{0\pi},m_{0\sigma},T)$ be the difference of the two functions on different sheets:
\begin{equation}
F(p^2,|\bm{p}|,m_{0\pi},m_{0\sigma},T)=I_{unphys}^{(2)}(p^2,|\bm{p}|,m_{0\pi},m_{0\sigma},T)-I^{(2)}(p^2,|\bm{p}|,m_{0\pi},m_{0\sigma},T).
\end{equation}
On the cut $0<p^2<(m_{0\sigma}-m_{0\pi})^2$, $I^{(2)}(p,m_{0\pi},m_{0\sigma},T)$ approaching from the above and $I_{unphys}^{(2)}(p,m_{0\pi},m_{0\sigma},T)$ approaching from the below are continuous:
\begin{equation}
I^{(2)}(p^2+i\eta,|\bm{p}|,m_{0\pi},m_{0\sigma},T)=I_{unphys}^{(2)}(p^2-i\eta,|\bm{p}|,m_{0\pi},m_{0\sigma},T),
\end{equation}
and therefore,
\begin{eqnarray}
F(p^2,|\bm{p}|,m_{0\pi},m_{0\sigma},T)
    &=&I^{(2)}(p^2+i\eta,|\bm{p}|,m_{0\pi},m_{0\sigma},T)-I^{(2)}(p^2-i\eta,|\bm{p}|,m_{0\pi},m_{0\sigma},T)\nonumber\\
    &=&\textrm{Disc}I^{(2)}(p^2,|\bm{p}|,m_{0\pi},m_{0\sigma},T)\nonumber\\
    &=&2i\textrm{Im}I^{(2)}(p^2,|\bm{p}|,m_{0\pi},m_{0\sigma},T)\nonumber\\
    &=&\frac{i}{8\pi\beta|\bm{p}|}\ln\frac{(1-\Exp{-\beta{\omega}_{\pi}^{+}})(1-\Exp{-\beta{\omega}_{\sigma}^{-}})}{(1-\Exp{-\beta{\omega}_{\pi}^{-}})(1-\Exp{-\beta{\omega}_{\sigma}^{+}})}.
\end{eqnarray}
where
\begin{equation}
\omega_{\pi}^{\pm}=-a_{\pi}\sqrt{p^2+|\bm{p}|^2}\pm c|\bm{p}|,\hspace{2.1cm}
\omega_{\sigma}^{\pm}=a_{\sigma}\sqrt{p^2+|\bm{p}|^2}\pm c|\bm{p}|,\nonumber\\
\end{equation}
which can be trivially extended for complex $p^2$.
Thus, the complete expression for $I_{unphys}^{(2)}(p^2,|\bm{p}|,m_{0\pi},m_{0\sigma},T)$ reads
\begin{eqnarray}
I_{unphys}^{(2)}(p^2,|\bm{p}|,m_{0\pi},m_{0\sigma},T) &=& \frac{1}{16\pi^2}[\frac{1}{\bar{\epsilon}}+2-a_{\pi}\ln(\frac{m_{0\pi}^2}{\kappa^2})-a_{\sigma}\ln(\frac{m_{0\sigma}^2}{\kappa^2}) - c\ln\frac{(c+a_{\pi})(c+a_{\sigma})}{(c-a_{\pi})(c-a_{\sigma})}]\nonumber\\
&&-\int^{\infty}_0\frac{dkk}{16\pi^2|\bm{p}|}
                           \left(\frac{n(\omega_{\pi k})}{\omega_{\pi k}}\ln\frac{((a_{\pi}p^2+k|\bm{p}|)^2-p_0^2\omega_{\pi k}^2)}{((a_{\pi}p^2-k|\bm{p}|)^2-p_0^2\omega_{\pi k}^2)}\right.  \nonumber \\
                       &&   \left. +\frac{n(\omega_{\sigma k})}{\omega_{\sigma k}}\ln\frac{((a_{\sigma}p^2+k|\bm{p}|)^2-p_0^2\omega_{\sigma k}^2)}{((a_{\sigma}p^2-k|\bm{p}|)^2-p_0^2\omega_{\sigma k}^2)}\right)\nonumber\\
&&+\frac{i}{8\pi\beta|\bm{p}|}\ln\frac{(1-\Exp{-\beta{\omega}_{\pi}^{+}})(1-\Exp{-\beta{\omega}_{\sigma}^{-}})}{(1-\Exp{-\beta{\omega}_{\pi}^{-}})(1-\Exp{-\beta{\omega}_{\sigma}^{+}})}.
\end{eqnarray}

\section{Imaginary Part of the self-energy}
The momentum dependence of the spectral function is caused by the self-energy $\Pi(p,T)$.
Since the real part of the self-energy is related to its imaginary part by the dispersion relation, the origin of the momentum dependence of the spectral function can be traced back to the imaginary part of the self-energy.
Therefore, we discuss here the imaginary part of the self-energy \cite{Weldon:1983jn}. We first consider the self-energy for the pion field:
\begin{eqnarray}
\lefteqn{\textrm{Im}\Pi_{\pi}(p,T)}\nonumber\\
&=&-\textrm{sign}(p_0)\frac{\lambda^2\xi^2}{9}\int\frac{dk^3}{(2\pi)^3}\frac{\pi}{4\omega_{\pi}\omega_{\sigma}}\Big\{[(1+n_{\pi})(1+n_{\sigma})-n_{\pi}n_{\sigma}]\delta(p_0-\omega_{\pi}-\omega_{\sigma}) \nonumber\\
&&\hspace{3cm}+[(1+n_{\sigma})n_{\pi}-(1+n_{\pi})n_{\sigma}]\delta(p_0+\omega_{\pi}-\omega_{\sigma}) \nonumber\\
&&\hspace{3cm}+[(1+n_{\pi})n_{\sigma}-(1+n_{\sigma})n_{\pi}]\delta(p_0-\omega_{\pi}+\omega_{\sigma}) \nonumber\\
&&\hspace{3cm}+[n_{\pi}n_{\sigma}-(1+n_{\pi})(1+n_{\sigma})]\delta(p_0+\omega_{\pi}+\omega_{\sigma}))\Big\},
\label{eq:imagselfenergypi}
\end{eqnarray}
where $n_{\pi}=n(\omega_{\pi})$, $n_{\sigma}=n(\omega_{\sigma})$, $\omega_{\pi}=\sqrt{\bm{k}^2+m_{0\pi}^2}$ and $\omega_{\sigma}=\sqrt{(\bm{p-k})^2+m_{0\sigma}^2}$.
The first term represents the probability for the process, $\tilde\pi \rightarrow \pi \sigma$, with the statistical weight $(1+n_{\pi})(1+n_{\sigma})$ minus the probability for the inverse process, $\pi\sigma \rightarrow \tilde\pi$, with the weight $n_{\pi}n_{\sigma}$, which is nonvanishing when $p^2>(m_{0\sigma}+m_{0\pi})^2$.
(When we describe processes we use the notation $\sigma$ and $\pi$ for on-shell particles while $\tilde\sigma$ and $\tilde\pi$ for off-shell particles associated with sigma and pion fields.)
The second term is due to the scattering of $\tilde\pi$ with particles in the heat bath and represents the probability for $\tilde\pi\pi \rightarrow \sigma$ with the weight $n_{\pi}(1+n_{\sigma})$ minus the probability for $\sigma \rightarrow \tilde\pi\pi$ with the weight $n_{\sigma}(1+n_{\pi})$.
It exists only at finite temperature.
If $m_{0\sigma}>m_{0\pi}$, the process, $\tilde\pi\pi \rightarrow \sigma$, is possible when $p^2<(m_{0\sigma}-m_{0\pi})^2$. 
The third term and the fourth term are anti-particle counter parts of the second term and the first term, respectively.

Imaginary part of the self-energy for sigma has a structure similar to that for pion, but the relevant processes for the former are $\tilde{\sigma}\rightarrow\pi\pi$ and $\tilde{\sigma}\rightarrow\sigma\sigma$ instead of $\tilde{\pi}\rightarrow\pi\sigma$ for the latter.
Hereafter we assume $p_0>0$.
Carrying out the integrations in Eq.(\ref{eq:imagselfenergypi}) we obtain
\begin{eqnarray}
    &&\textrm{Im}\Pi_{\pi}(p,T)\nonumber\\
    &=&\left\{
    \begin{array}{ll}
    \displaystyle{-\frac{\lambda^2\xi^2}{9}\frac{1}{16\pi|\bm{p}|} \int^{\omega_{\pi}^+}_{\omega_{\pi}^-}
    d\omega_{\pi}\left[1+n(\omega_{\pi})+n(p_0-\omega_{\pi})\right]} & p^2>(m_{0\sigma}+m_{0\pi})^{2}\\
    \displaystyle{-\frac{\lambda^2\xi^2}{9}\frac{1}{16\pi|\bm{p}|} \int^{\omega_{\pi}^+}_{\omega_{\pi}^-}d\omega_{\pi}\left[n(\omega_{\pi})-n(p_0+\omega_{\pi})\right]} & 0<p^2<(m_{0\sigma}-m_{0\pi})^{2}\\
    \displaystyle{-\frac{\lambda^2\xi^2}{9}\frac{1}{16\pi|\bm{p}|} \left\{\int^{\infty}_{\omega_{\pi}^-}d\omega_{\pi}\left[n(\omega_{\pi})-n(p_0+\omega_{\pi})\right]\right.}& \\\hspace{2.2cm}\displaystyle{\left. + \int^{\infty}_{\omega_{\sigma}^+}d\omega_{\sigma}\left[n(\omega_{\sigma})-n(p_0+\omega_{\sigma})\right]\right\}} & -(m_{0\sigma}^2-m_{0\pi}^{2})<p^2<0\\
    \displaystyle{-\frac{\lambda^2\xi^2}{9}\frac{1}{16\pi|\bm{p}|} \left\{\int^{\infty}_{\omega_{\pi}^-}d\omega_{\pi}\left[n(\omega_{\pi})-n(p_0+\omega_{\pi})\right]\right.}& \\\hspace{2.2cm}\displaystyle{\left. + \int^{\infty}_{\omega_{\sigma}^-}d\omega_{\sigma}\left[n(\omega_{\sigma})-n(p_0+\omega_{\sigma})\right]\right\}} & p^2<-(m_{0\sigma}^2-m_{0\pi}^{2})\\
    \end{array}\right.\nonumber\\
    &=&\left\{
    \begin{array}{ll}
    \displaystyle{-\frac{\lambda^2\xi^2}{9}\left[\frac{1}{8\pi}c+\frac{1}{16\pi\beta|\bm{p}|}\ln\frac{(1-\Exp{-\beta\omega_{\pi}^+})(1-\Exp{-\beta{\omega}_{\sigma}^+})}{(1-\Exp{-\beta{\omega}_{\pi}^-})(1-\Exp{-\beta{\omega}_{\sigma}^-})}\right]} & p^2>(m_{0\sigma}+m_{0\pi})^{2}\\
    \displaystyle{-\frac{\lambda^2\xi^2}{9}\frac{1}{16\pi\beta|\bm{p}|}\ln\frac{(1-\Exp{-\beta\omega_{\pi}^+})(1-\Exp{-\beta{\omega}_{\sigma}^-})}{(1-\Exp{-\beta{\omega}_{\pi}^-})(1-\Exp{-\beta{\omega}_{\sigma}^+})}} & \hspace{-1.7cm} -(m_{0\sigma}^2-m_{0\pi}^{2})<p^2<(m_{0\sigma}-m_{0\pi})^{2}\\
    \displaystyle{-\frac{\lambda^2\xi^2}{9}\frac{1}{16\pi\beta|\bm{p}|}\ln\frac{(1-\Exp{-\beta\omega_{\pi}^+})(1-\Exp{-\beta{\omega}_{\sigma}^+})}{(1-\Exp{-\beta{\omega}_{\pi}^-})(1-\Exp{-\beta{\omega}_{\sigma}^-})}} & p^2<-(m_{0\sigma}^2-m_{0\pi}^{2}),\\
    \end{array}\right.\nonumber\\
    \label{eq:Discselfenergy}
    \end{eqnarray}
where $\omega_{\pi}^{\pm}$ and $\omega_{\sigma}^{\pm}$  are
\begin{equation}
\omega_{\pi}^{\pm}=||a_{\pi}p_0|\pm c|\bm{p}||,\hspace{2.1cm}
\omega_{\sigma}^{\pm}=||a_{\sigma}p_0|\pm c|\bm{p}||,
\end{equation}
and, $a_{\pi}$, $a_{\sigma}$ and c are shown in Eq.(\ref{eq:parameter}).

We first discuss the behaviour of the imaginary part at large momentum.
As $|\bm{p}|\rightarrow \infty$, the temperature-dependent part of the imaginary part behaves as
\begin{equation}
  \textrm{Im}\Pi_{\pi}(p,T)\rightarrow
-\frac{\lambda^2\xi^2}{9}\frac{1}{16\pi\beta|\bm{p}|}\Exp{-\beta\omega_{\pi}^-}.
\end{equation}
This can be understood as follows.
As $|\bm{p}|\rightarrow\infty$, $\omega_{\pi}^{\pm}\rightarrow|a_{\pi}\pm c||\bm{p}|$ and $\omega_{\sigma}^{\pm}\rightarrow|a_{\sigma}\pm c||\bm{p}|$ therefore the energies of $\pi$ and $\sigma$, which participate in the process, become also large.
Since the temperature-dependent imaginary part of the self-energy includes a Bose distribution factor for either $\pi$ or $\sigma$, which vanishes at the large energy limit, the temperature-dependent imaginary part of the self-energy also vanishes when $|\bm{p}|\rightarrow\infty$.
This implies that thermal effects effectively decrease at large momentum.

It should be noted, however, that the temperature dependence of the self-energy does not completely vanish at large momentum: the contribution of the thermal tad pole diagrams to the real part of the self-energy and the temperature dependence of the mass parameter due to the optimization procedure do not vanish at large momentum.

We next ask at which momentum the imaginary part of self-energy becomes maximum.
The process which generates the imaginary part is $\tilde\pi\pi\rightarrow\sigma$.
At low temperature, the distribution function $n(\omega)$ rapidly damps as $|\bm{p}|$ increases, i.e. the heat bath is occupied by pions almost at rest, while $\tilde\pi$ can interact with only pions whose energy lies between $\omega_{\pi}^-$ and $\omega_{\pi}^+$.
This means that the integral in Eq.(\ref{eq:Discselfenergy}) has a sharp peak at the momentum $|\bm{p}|$ which satisfies $\omega_{\pi}=m_{0\pi}$, i.e. $|\bm{p}|=cp^2/m_{0\pi}$.
This momentum maximizes the imaginary part of the self-energy, which is proportional to the integral times the kinematical factor $1/|\bm{p}|$.
This is because around $|\bm{p}|=0$ the integral is proportional to $|\bm{p}|$ and cancels the kinematical factor, which makes the imaginary part of the self-energy constant, while around $|\bm{p}|=cp^2/m_{0\pi}$ the integral has a sharp peak and the dependence of the kinematical factor can be neglected.
When the temperature is not low, the momentum which maximizes the imaginary part cannot be expressed in such a simple form and is determined by balancing the integral and the kinematical factor $1/|\bm{p}|$.

A comment on higher loop effects is in order here.
Let us consider the process $\tilde{\pi}\rightarrow\pi\sigma$.
In Eq.(\ref{eq:Discselfenergy}) the term $\frac{\lambda^2\xi^2}{9}$ is nothing but $TT^*$ where $T$ is the tree-level T-matrix for the process, $\tilde{\pi}\rightarrow\pi\sigma$.
It is speculated that one can take account of higher loop effects by replacing the tree-level T-matrix by the one including higher loop effects.
If this is the case, the above argument on the momentum dependence of the imaginary part of the self-energy holds beyond the one-loop order as far as the T-matrix is not singular at large momentum.
Of course, new processes such as $\tilde{\pi}\sigma\rightarrow\pi\sigma$ exist in higher loop terms, which change the structure of the spectral function.
This is a future problem.

\section{Numerical Results}

\subsection{$T$ dependence for $\bm{p}=0$}
In this subsection we discuss how the spectral functions and the poles of the propagators depend on the temperature with the spatial momentum kept fixed ($\bm{p}=0$).
\subsubsection{$\pi$ channel}
%
Fig.\ref{Tdependencepi} shows the temperature dependence in the $\pi$ channel.
When $T=0$ the propagator has a pole, (A), at $p^2=(145{\textrm{MeV}})^2$ on the physical sheet of the complex $p^2$ plane which shows up as a $\delta$-function peak in the spectral function.
Another pole, (B), exists at $p^2=(447\textrm{MeV})^2$ on the unphysical sheet which is too far from the physical region to have a visible effect on the spectral function.
As the temperature increases, the pole (A) moves down and away from the real axis and the peak of the spectral function becomes broader. The pole (B) moves left on the real axis and approaches the $\tilde{\pi}\pi \rightarrow\sigma$ threshold as the temperature approaches $T=165$MeV which causes the enhancement of the threshold. When $T=165$MeV, the pole crosses the $\tilde{\pi}\pi \rightarrow\sigma$ threshold at $p^2=(m_{0\sigma}-m_{0\pi})^2$ and appears on the physical sheet. When $T>165\textrm{MeV}$ a bound state peak exists.

\subsubsection{$\sigma$ channel}
%
Fig.\ref{Tdependencesigma} shows the temperature dependence in the $\sigma$ channel.
When $T=0$ the propagator has a pole, (C), at $p^2=(556 \textrm{MeV})^2+i(368\textrm{MeV})^2$ on the unphysical sheet of the complex $p^2$ plane which shows up as a broad bump around $p^2=(550\textrm{MeV})^2$ in the spectral function.
Another pole, (D), exists at $p^2=(207\textrm{MeV})^2$ on the unphysical sheet of the complex $p^2$ plane which seems to provide a little shoulder at the $\tilde{\sigma}\rightarrow\pi\pi$ threshold.
As the temperature increases the pole (C) moves left and a little bit down and the bump becomes broader and shifts towards lower $p^2$.
The pole (D) moves right on the real axis and approaches the $\tilde{\sigma}\rightarrow\pi\pi$ threshold as the temperature approaches $145 \textrm{MeV}$ which causes the enhancement of the spectral function at the $\tilde{\sigma}\rightarrow\pi\pi$ threshold.
When $T=145\textrm{MeV}$, the pole crosses the $\tilde{\sigma}\rightarrow\pi\pi$ threshold  at $p^2=4m_{0\pi}^2$ and appears on the physical sheet. When $T>145\textrm{MeV}$ a bound state peak exists.

A comment is in order here on the threshold enhancement of the spectral function and the behaviour of the poles of the propagator.
What is observed above is nothing but the well-known phenomena in the quantum mechanical s-wave scattering \cite{Newton:1982qc}.

Now, we would like to compare our results for $\bm{p}=0$ with previous results.
In ref.\cite{Chiku:1998kd} Chiku and Hatsuda calculated the spectral functions in the $\pi$ and the $\sigma$ channels using the OPT.
We just reproduced their results for the spectral functions.
We also studied the pole of the correlation function in the complex $p^2$ plane, which was not done in ref.\cite{Chiku:1998kd}.
In ref.\cite{Patkos:2002xb}, Patkos et al. applied large N expansion, and obtained results similar to ref.\cite{Chiku:1998kd}.
Then, in ref.\cite{Patkos:2002vr} they analyzed the relationship between the scalar-isoscalar spectral function and the second Riemann sheet pole in the same channel.
The behaviour of the pole around the $2\pi$ threshold is similar to the pole, D, in Fig.\ref{Tdependencesigma}.
A crucial difference is that in the present paper the pole, which is responsible for the threshold enhancement of the spectral function in the sigma channel, is different from the one for the peak at zero temperature, while in ref.\cite{Patkos:2002vr} the former is identified to have continuously moved from the latter as the temperature increases.
Which scenario is closer to reality should be clarified in future.
Also pointed out in ref.\cite{Patkos:2002xb,Patkos:2002vr} was that the range of the validity of the model is severely restricted by the unavoidable presence of a tachyonic pole.
In our approximate solution, however, there is no tachyonic pole within discussed region of the parameters.
Probably, the problem of a tachyonic pole will arise when higher loop terms are taken into account.
In ref.\cite{Dobado:2002xf}, Dobado et al. studied the mass and the width of $\sigma$ and $\rho$ using chiral unitary approach.
They searched for the poles of the T-matrix of $\pi\pi$ scattering and showed that the width grows with temperature, while the mass decreases slightly.
The behaviour of the pole in the $\sigma$ channel in ref.\cite{Dobado:2002xf} is consistent with the behaviour of the pole C in Fig.\ref{Tdependencesigma}.
However, they did not discuss the the pole, which corresponds to the pole D in Fig.\ref{Tdependencesigma}.
Therefore, it is important if such a pole exists or not in their approach.

\subsection{$|\bm{p}|$ dependence}
In this subsection we discuss how the spectral functions and the poles of the propagators depend on the spatial momentum with the temperature fixed ($T=50$, $145$ and $170$ MeV).
\subsubsection{$\pi$ channel} 
Figs.\ref{pdependenceT=50pi}, \ref{pdependenceT=145pi} and \ref{pdependenceT=170pi} show the momentum dependence in the $\pi$ channel at $T=50$, $145$ and $170$ MeV, respectively.
When $T=50\textrm{MeV}$ the pole (A) moves down at the beginning, then to the right and then up again as the momentum increases, approaching the real axis as $|\bm{p}|\rightarrow \infty$. As a consequence, the peak becomes broader up to about $|\bm{p}|=500\textrm{MeV}$ and then becomes narrower again, approaching the $\delta$-function as $|\bm{p}|\rightarrow\infty$.
At higher temperatures, $T=145$ and $170$ MeV, similar momentum dependence is observed for the spectral function and the pole of the propagator, but as the temperature increases the variation of the spectral function and the movement of the pole becomes larger.
On the other hand, the pole (B) moves right and away from the $\tilde{\pi}\pi\rightarrow\sigma$ threshold as the momentum increase at $145$ MeV. As a consequence, the threshold enhancement of the spectral function decreases.
At $T=170$ MeV, the pole (B), which lies on the physical sheet for $|\bm{p}|=0$, crosses the threshold and moves into the unphysical sheet as $|\bm{p}|$ increases.
Accordingly, the bound state peak becomes the sharp threshold enhancement and the it decreases.

\subsubsection{$\sigma$ channel} 
Figs.\ref{pdependenceT=50sigma}, \ref{pdependenceT=145sigma} and \ref{pdependenceT=170sigma} show the momentum dependence in the $\sigma$ channel at $T=50$, $145$ and $170\textrm{MeV}$, respectively.
When the temperature is as small as $50\textrm{MeV}$ the momentum dependence of the spectral function is very small.
At high temperature $T=145$MeV, the pole (D), which is located near the $\tilde{\sigma}\rightarrow\pi\pi$ threshold for $|\bm{p}|=0$, moves left and away from the threshold and then threshold enhancement of the spectral function decreases.
At $T=170$MeV, the pole (D), which lies on the physical sheet for $|\bm{p}|=0$, crosses the threshold and moves into the unphysical sheet as $|\bm{p}|$ increases.
Accordingly, the bound state peak becomes the sharp threshold enhancement and the it decreases.

It should be noted that the numerical results presented in this section are consistent with the discussion in the previous section.
\section{Summary} 
In this paper we have studied the momentum dependence of the spectral function in the $\mathcal{O}(4)$ linear sigma model.

We analyzed the spatial momentum dependence of the imaginary part of the self-energy.
The temperature-dependent imaginary part of the self-energy was found to vanish in the large momentum limit.
This is because the energy of the particles in the heat bath which participate in the process becomes large and therefore the Bose distribution function vanishes.
However, the temperature-dependent part of the self-energy does not completely vanish at large momentum: the contribution of the thermal tad pole diagrams to the real part of the self-energy and the temperature dependence of the mass parameter due to the optimization procedure do not vanish at large momentum.

We then calculated the spectral functions and searched for the poles of the propagators.
First, we discussed the temperature dependence for zero spatial momentum.
We reproduced the spectral functions in both $\pi$ and $\sigma$ channels in the previous work and found that the pole, which is responsible for the threshold enhancement of the spectral function in both $\pi$ and $\sigma$ channels, is different from the one for the peak at zero temperature.
Secondly, we discussed the spatial-momentum dependence.
When the temperature is low the spatial momentum dependence is also small.
As the temperature becomes higher the spatial momentum dependence becomes also large.
When the spatial momentum is smaller than the temperature, the spectral functions do not deviate much from those at zero spatial momentum.
Once the spatial momentum becomes comparable with the temperature, the deviation becomes considerable.
As the momentum becomes much larger than the temperature, thermal effects effectively decrease.

These observations have an implication for heavy-ion collision experiments:
When one extracts information about the change of intrinsic properties of hadrons at finite temperature from their spectral functions measured in high energy heavy-ion collision experiments, one should keep in mind spatial momentum dependence discussed in the present paper.

We expect that the essential part of the results in the present paper holds for other hadrons, since it is based basically on the kinematics of the particles under consideration and thermal particles in the heat bath.

Finally, a comment on the higher-loop effects is in order.
At finite temperature there exist various physical processes which are not taken into account in the present one-loop calculation.
If we include those processes in the calculation, the structure of the spectral functions might be significantly modified by the following reason:
The sigma meson at zero temperature has large width due to the strong coupling with two pions.
It was found in the one-loop calculation that the spectrum of $\sigma$
at finite temperature is enhanced near the $\tilde{\sigma}\rightarrow\pi\pi$ threshold.
This is because as the temperature increases the mass of $\sigma$ decreases while that of $\pi$ increases.
However, what is not taken into account in the one-loop calculation is the effect of the thermal width of $\pi$ in the process $\tilde{\sigma}\rightarrow\pi\pi$.
This is taken into account in the calculation to two-loop order.
If this effect is included this effect the sharp threshold does not exist any more and therefore the enhancement might be smeared out.
The extension of the calculation to two-loop order is now in progress 
\footnote{T. Nishikawa and O. Morimatsu and Y. Hidaka, in preparation.}. 

\appendix* 
\section{One-Loop self-energy At Finite Temperature}  
   \begin{figure}[htbp]
    \begin{center}
    \includegraphics[width=1\linewidth,clip]{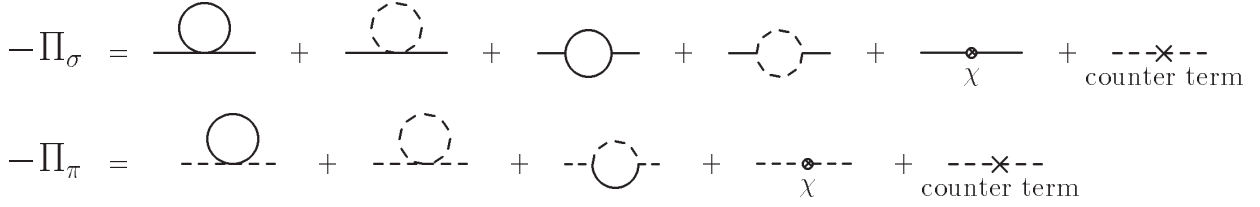}
        \caption{Feynman diagrams representing one-loop self-energy for $\pi$ and $\sigma$. Solid and dashed lines correspond to $\sigma$ and $\pi$ respectively.}
        \label{fig:one-loop}
    \end{center}   
    \end{figure} 
    Fig.\ref{fig:one-loop} read
       \begin{eqnarray}
    {\Pi}_{\sigma}(p)
    &=& \lambda(\frac{1}{2}I^{(1)}(m_{0\pi},T)+\frac{1}{2}I^{(1)}(m_{0\sigma},T))
      -\lambda^2\xi^2(\frac{1}{2}I^{(2)}(p,m_{0\sigma},m_{0\sigma},T)\nonumber\\ &&+\frac{1}{6}I^{(2)}(p,m_{0\pi},m_{0\pi},T))-(m^2-\mu^2)+\textrm{(counter terms)},\\
    {\Pi}_{\pi}(p) 
    &=& \lambda(\frac{5}{6}I^{(1)}(m_{0\pi},T)+\frac{1}{6}I^{(1)}(m_{0\sigma},T))-\lambda^2\xi^2\frac{1}{9}I^{(2)}(p,m_{0\pi},m_{0\sigma},T)\nonumber\\
    &&-(m^2-\mu^2)+\textrm{(counter terms)},
    \end{eqnarray}
   \begin{eqnarray}
   I^{(1)}(m_1,T)         &=& T\sum_m\int\frac{dk^3}{(2\pi)^3}\frac{-1}{k^2-m_1^2},  \\
   I^{(2)}(p,m_1,m_2,T) &=& T\sum_m\int\frac{dk^3}{(2\pi)^3}\frac{-1}{k^2-m_1^2}\frac{-1}{(k-p)^2-m_2^2},
   \label{eq:I1I2_}
   \end{eqnarray}
   where $p=(i\omega_n,\bm{p})$,    $p^2=(i\omega_n)^2-\bm{p}^2$.\\
   Summing up Matsubara frequencies in Eq.(\ref{eq:I1I2_}), $I^{(1)}$ and $I^{(2)}$ can be separated temperature-independent part and temperature-dependent part:
    \begin{eqnarray}
 I^{(1)}(m_1,T)&=& I^{(1)}_0(m_1)+I^{(1)}_T(m_1,T),\\
 I^{(2)}(p,m_1,m_2,T)&=&I^{(2)}_0(p,m_1,m_2)+I^{(2)}_T(p,m_1,m_2,T),\\
 I^{(1)}_0(m_1,T)&=&\int\frac{d^{3}k}{(2\pi)^{3}}\frac{1}{2\omega_1},\\
 I^{(1)}_T(m_1,T)&=&\int\frac{d^{3}k}{(2\pi)^{3}}\frac{n(\omega_1)}{\omega_1},\\
   I^{(2)}_0(p,m_1,m_2,T)&=&\int\frac{d^{3}{k}}{(2\pi)^{3}}\frac{-1}{4\omega_1\omega_2}\Big(\frac{1}{i\omega_n-\omega_1-\omega_2}-\frac{1}{i\omega_n+\omega_1+\omega_2}\Big),\\
    I^{(2)}_T(p,m_1,m_2,T)&=&\int\frac{d^{3}{k}}{(2\pi)^{3}}\frac{-1}{4\omega_1\omega_2}\Big\{[n(\omega_1)+n(\omega_2)](\frac{1}{i\omega_n-\omega_1-\omega_2}-\frac{1}{i\omega_n+\omega_1+\omega_2})\nonumber\\
    &&\textrm{\hspace{0.7cm}}+[n(\omega_1)-n(\omega_2)](\frac{1}{i\omega_n+\omega_1-\omega_2}-\frac{1}{i\omega_n-\omega_1+\omega_2})\Big\},
    \label{I1I2_2}
   \end{eqnarray}
   where $\omega_1=\sqrt{\bm{k}^2+m_{1}^2}$, $\omega_2=\sqrt{(\bm{p}-\bm{k})^2+m_{2}^2}$, $n(\omega)=\frac{1}{\Exp{\beta\omega}-1}$. Performing analytic continuation $i\omega_n\rightarrow p_0(1+i\eta)$ $(\eta\rightarrow 0^+)$ and carrying out the integration in Eq.(\ref{I1I2_2}), we obtain
\begin{eqnarray}
   I_0^{(1)}(m_1)         &=& -\frac{m_1^2}{16\pi^2}(\frac{1}{\bar{\epsilon}}+1-\ln(\frac{m_1^2}{\kappa^2})), \\
   I_0^{(2)}(p,m_1,m_2) &=& \frac{1}{16\pi^2}[\frac{1}{\bar{\epsilon}}+2-a_{1}\ln(\frac{m_1^2}{\kappa^2})-a_{2}\ln(\frac{m_2^2}{\kappa^2})  \nonumber\\
                      &&  - \left\{
   \begin{array}{l}
   c(\ln\frac{|c+a_{1}||c+a_{2}|}{|c-a_{1}||c-a_{2}|}-2i\pi)] \\  
                   \textrm{\hspace{2cm}for } (m_1+m_2)^2<p^2 \\
   2c(\arctan\frac{a_{1}}{c}+\arctan\frac{a_{2}}{c})] \\  
                   \textrm{\hspace{2cm}for } (m_1-m_2)^2<p^2<(m_1+m_2)^2 \\
   c\ln\frac{|c+a_{1}||c+a_{2}|}{|c-a_{1}||c-a_{2}|}]\\
                   \textrm{\hspace{2cm}for } p^2<(m_1-m_2)^2 ,
   \end{array}
   \right. \\
   I_0^{(2)}(p,m_1,m_1)     &=& \frac{1}{16\pi^2}[\frac{1}{\bar{\epsilon}}+2-\ln(\frac{m_1^2}{\kappa^2})\nonumber\\ 
                      &&\textrm{    } -\left\{
   \begin{array}{ll}
   c_2(2\ln\frac{|2c_2+1|}{|2c_2-1|}-2i\pi)] &\textrm{ for } p^2>4m^2 \\
   4c_2\arctan\frac{1}{2c_2}]            &\textrm{ for } 0<p^2<4m^2 \\
   c_2(2\ln\frac{|2c_2+1|}{|2c_2-1|})]       &\textrm{ for } p^2<0, \\
   \end{array}
   \right.
   \end{eqnarray}
   Here $\bar{\epsilon}^{-1}=\frac{2}{4-d}-\gamma+\ln4\pi,\gamma$ is Euler constant, $\kappa$ is the renormalization point, and
    \begin{eqnarray}
    a_{1}&=&\frac{1}{2}\left\{1-\frac{m_2^2-m_1^2}{p^2}\right\},\hspace{1cm}
a_{2}=\frac{1}{2}\left\{1-\frac{m_1^2-m_2^2}{p^2}\right\},\nonumber\\
    c&=&\frac{1}{2}\sqrt{\left|\left(1-\frac{(m_1+m_2)^2}{p^2}\right)\left(1-\frac{(m_1-m_2)^2}{p^2}\right)\right|} \textrm{,\hspace{0.6cm}}c_2=\frac{1}{2}\sqrt{\left|\frac{p^2-4m_1^2}{p^2}\right|.}
    \end{eqnarray}
    Temperature-dependent part are
    \begin{eqnarray}
    I_T^{(1)}(m_1,T)       &=& \int^{\infty}_0\frac{dkk^2}{2\pi^2}\frac{n(\omega_{1k})}{\omega_{1k}},  \\
    I_T^{(2)}(p,m_1,m_2,T) &=& -\int^{\infty}_0\frac{dkk}{16\pi^2|\bm{p}|}
                           \left(\frac{n(\omega_{1k})}{\omega_{1k}}\ln\frac{|(k|\bm{p}|+a_{1}p^2)^2-p_0^2\omega_{1k}^2|}{|(k|\bm{p}|-a_{1}p^2)^2-p_0^2\omega_{1k}^2|}\right.  \nonumber \\
                       &&   \left. +\frac{n(\omega_{2k})}{\omega_{2k}}\ln\frac{|(k|\bm{p}|+a_{2}p^2)^2-p_0^2\omega_{2k}^2|}{|(k|\bm{p}|-a_{2}p^2)^2-p_0^2\omega_{2k}^2|}\right)  \nonumber\\
                       && +\left\{
    \begin{array}{l}
    \frac{i}{16\pi\beta|\bm{p}|}\ln\frac{
    (1-\Exp{-\beta{\omega}_{1}^{+}})(1-\Exp{-\beta{\omega}_{2}^{+}})}{(1-\Exp{-\beta{\omega}_{1}^{-}})(1-\Exp{-\beta{\omega}_{2}^{-}})}\\  
                            \textrm{\hspace{1cm}for } p^2>(m_1+m_2)^2,p^2<-|m_1^2-m_2^2|  \\
    \frac{i}{16\pi\beta|\bm{p}|}\textrm{sign}(m_2-m_1)\ln\frac{(1-\Exp{-\beta{\omega}_{1}^{+}})(1-\Exp{-\beta{\omega}_{2}^{-}})}{(1-\Exp{-\beta{\omega}_{1}^{-}})(1-\Exp{-\beta{\omega}_{2}^{+}})} \\ 
                            \textrm{\hspace{1cm}for } -|m_1^2-m_2^2|<p^2<(m_1-m_2)^2 \\
    0                               \\  
                            \textrm{\hspace{1cm}for } (m_1-m_2)^2<p^2<(m_1+m_2)^2, 
    \end{array}
    \right.  \\
    I_T^{(2)}(p,m_1,m_1,T)&=&\frac{-1}{8\pi^2|\bm{p}|}\int^{\infty}_0dkk\frac{n(\omega_{1k})}{\omega_{1k}}\ln\frac{|(k|\bm{p|}+\frac{p^2}{2})^2-p_0^2\omega_{1k}^2|}{|(k|\bm{p|}-\frac{p^2}{2})^2-p_0^2\omega_{1k}^2|}   \nonumber\\
    && +\left\{
    \begin{array}{ll}
    \frac{i}{8\pi\beta|\bm{p}|}\ln\frac{1-\exp(-\beta|\frac{1}{2}|p_0|+|\bm{p}|c_2|)}{1-\exp(-\beta|\frac{1}{2}|p_0|-|\bm{p}|c_2|)} & \textrm{ for }p^2>4m_1^2 \textrm{ , }p^2<0 \\
          0 & \textrm{ for }0<p^2<4m_1^2,
    \end{array}
    \right.
    \end{eqnarray}
    where
    \begin{eqnarray}
    {\omega}_{ik}&=&\sqrt{k^2+m_i^2}, \nonumber\\
    {\omega}_{i}^{\pm}&=&\left||p_0a_{i}|\pm|\bm{p}|c\right|.
    \end{eqnarray}

%
\begin{figure}
    \includegraphics[width=.9\linewidth,clip]{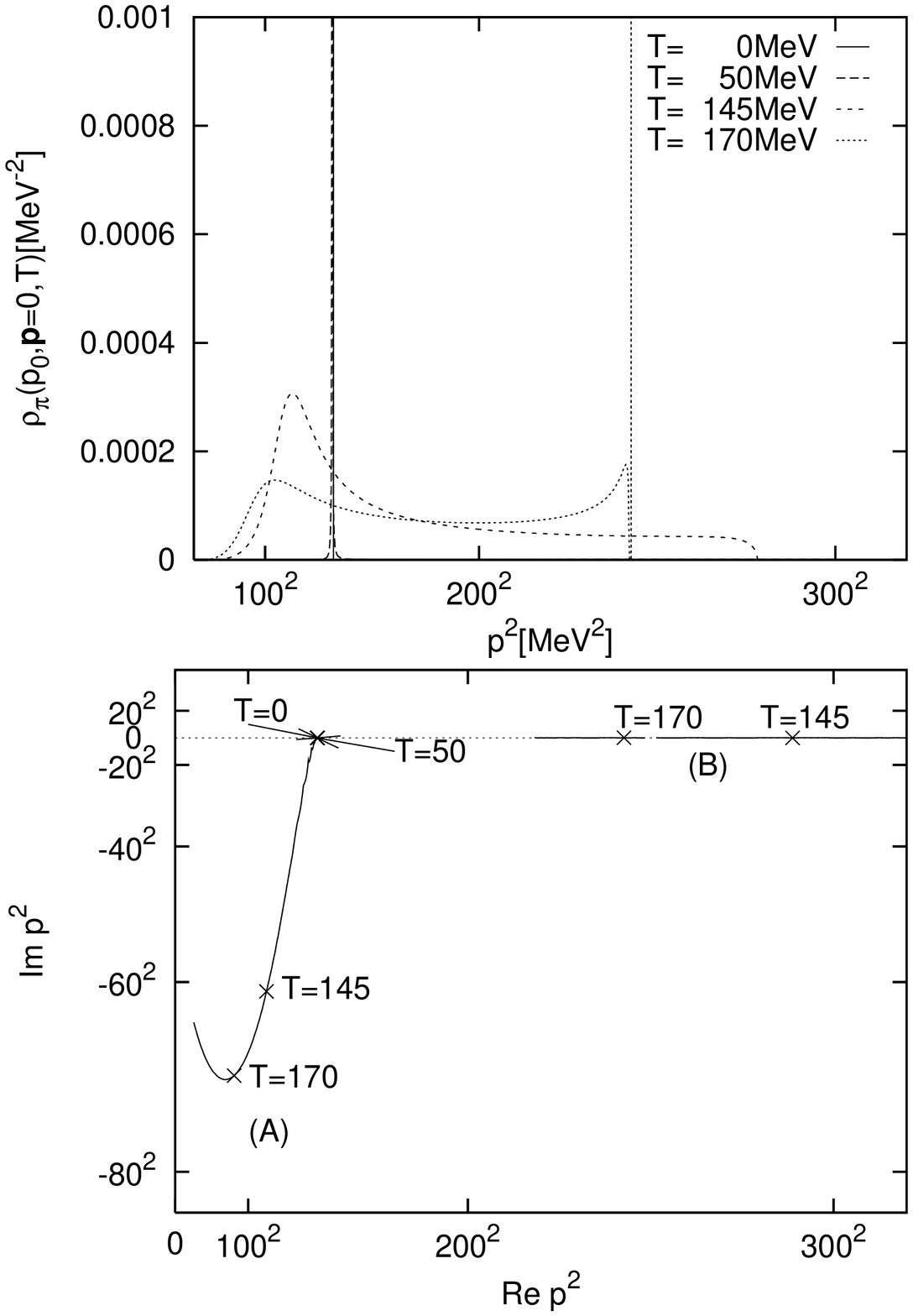}\\
    \caption{Spectral function and the pole position of propagator at $T=0, 50, 145, $ and $ 170$ MeV for $|\bm{p}|$=0 in the $\pi$ channel.}
    \label{Tdependencepi}
\end{figure}
\begin{figure}
    \includegraphics[width=.9\linewidth,clip]{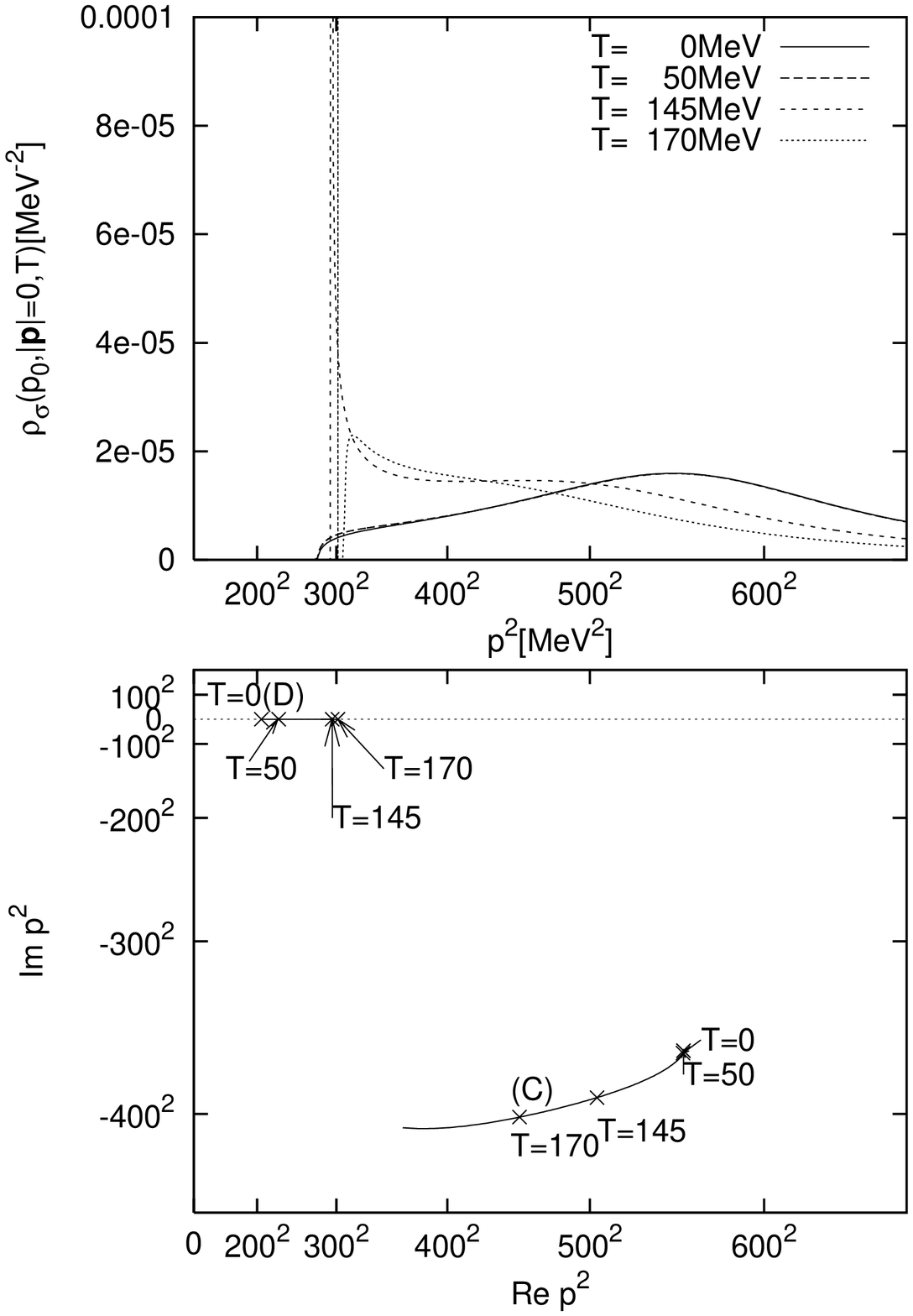}\\
    \caption{Spectral function and the pole position of propagator at $T=0, 50, 145, $ and $ 170$ MeV for $|\bm{p}|$=0 in the $\sigma$ channel.}
    \label{Tdependencesigma}
\end{figure}
\begin{figure}
    \includegraphics[width=.9\linewidth,clip]{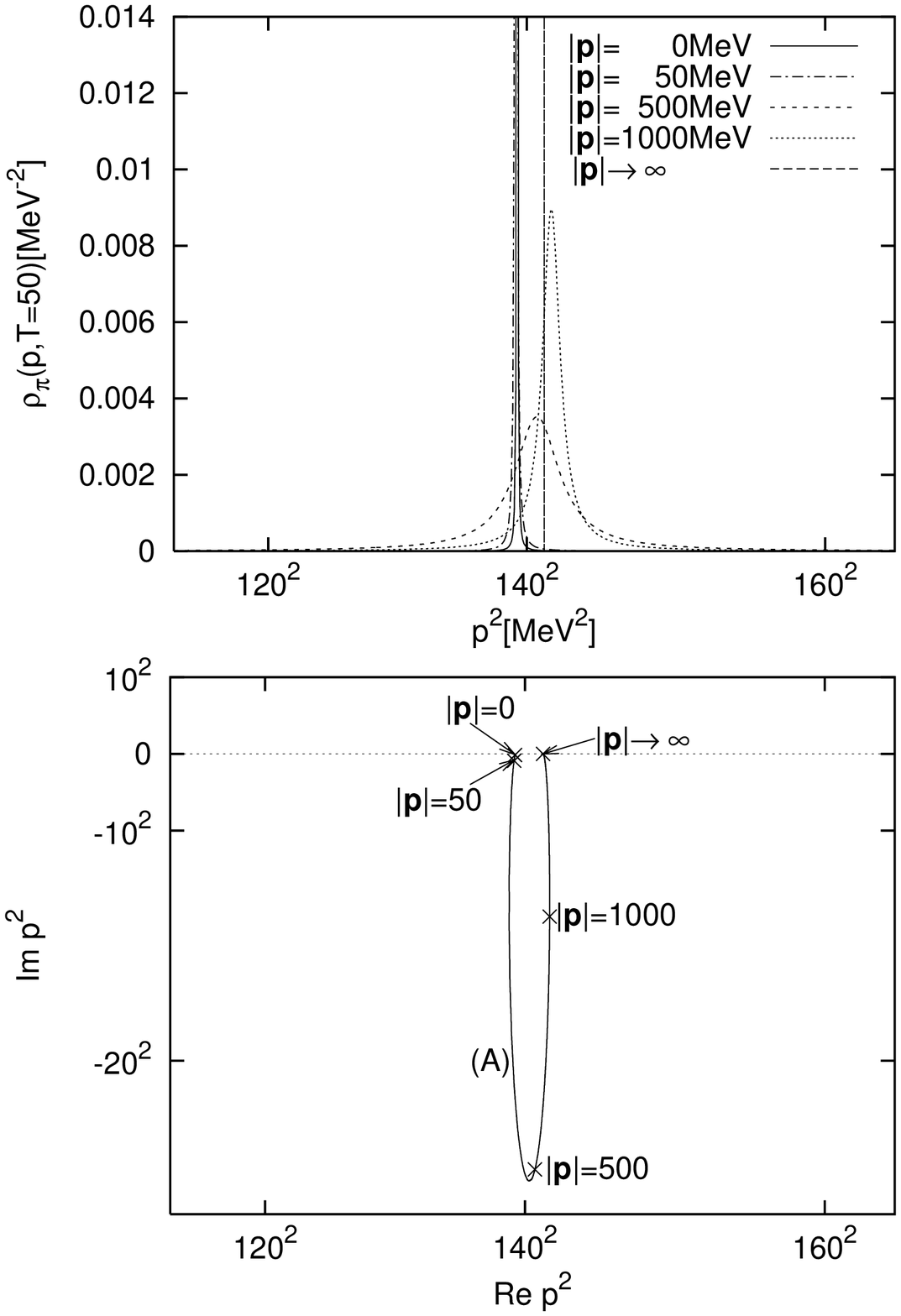} 
    \caption{Spectral function and the pole position of propagator at $T=50$MeV for $|\bm{p}|$=0, 50, 500, 1000, and $|\bm{p}|\rightarrow \infty$ in the $\pi$ channel.}
    \label{pdependenceT=50pi}
\end{figure}
\begin{figure}[p]
    \includegraphics[width=.9\linewidth,clip]{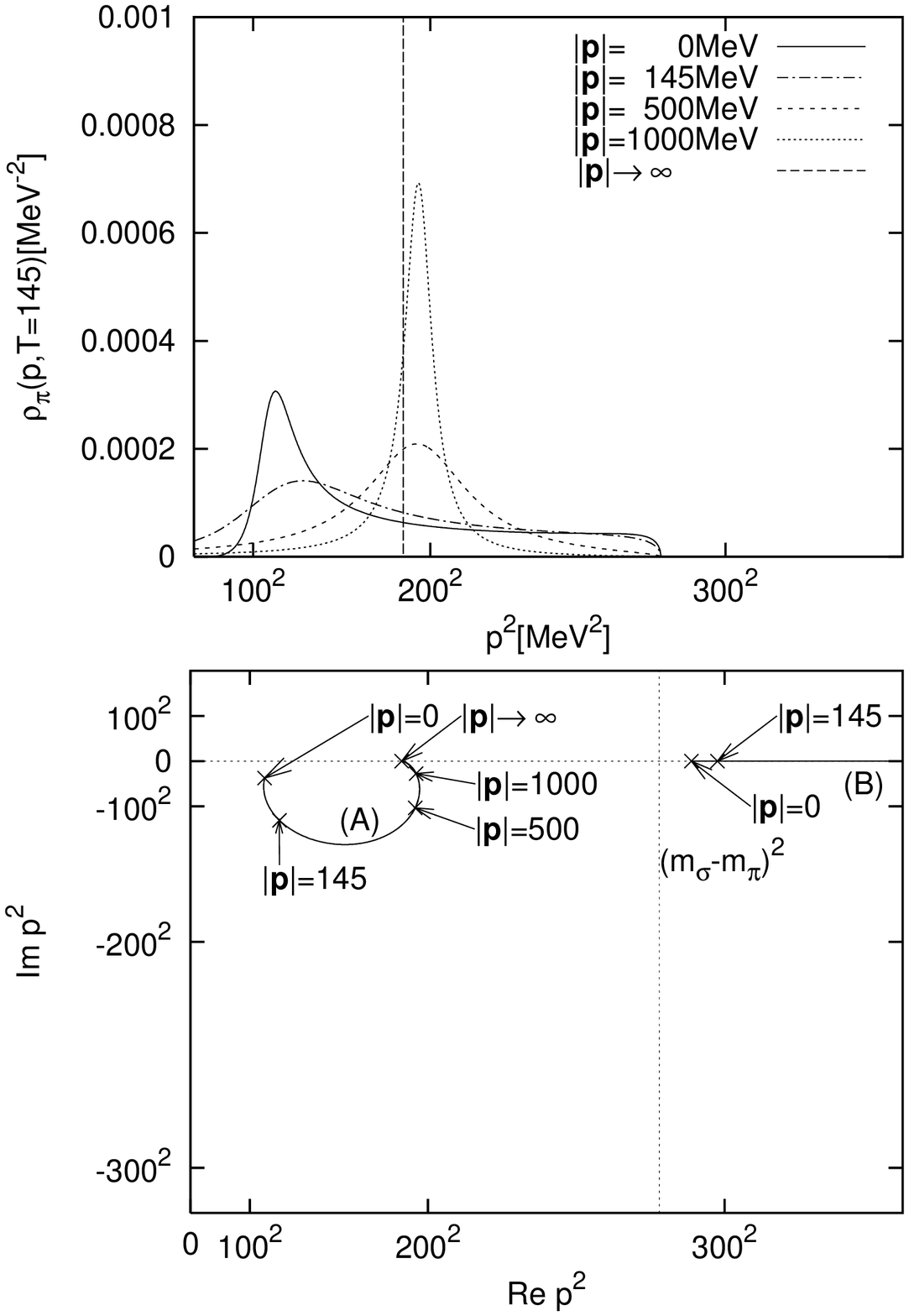} 
    \caption{Spectral function and the pole position of propagator at $T=145$MeV for $|\bm{p}|$=0, 145, 500, 1000, and $|\bm{p}|\rightarrow \infty$ in the $\pi$ channel.}
    \label{pdependenceT=145pi}
\end{figure}
\begin{figure}
    \includegraphics[width=.9\linewidth,clip]{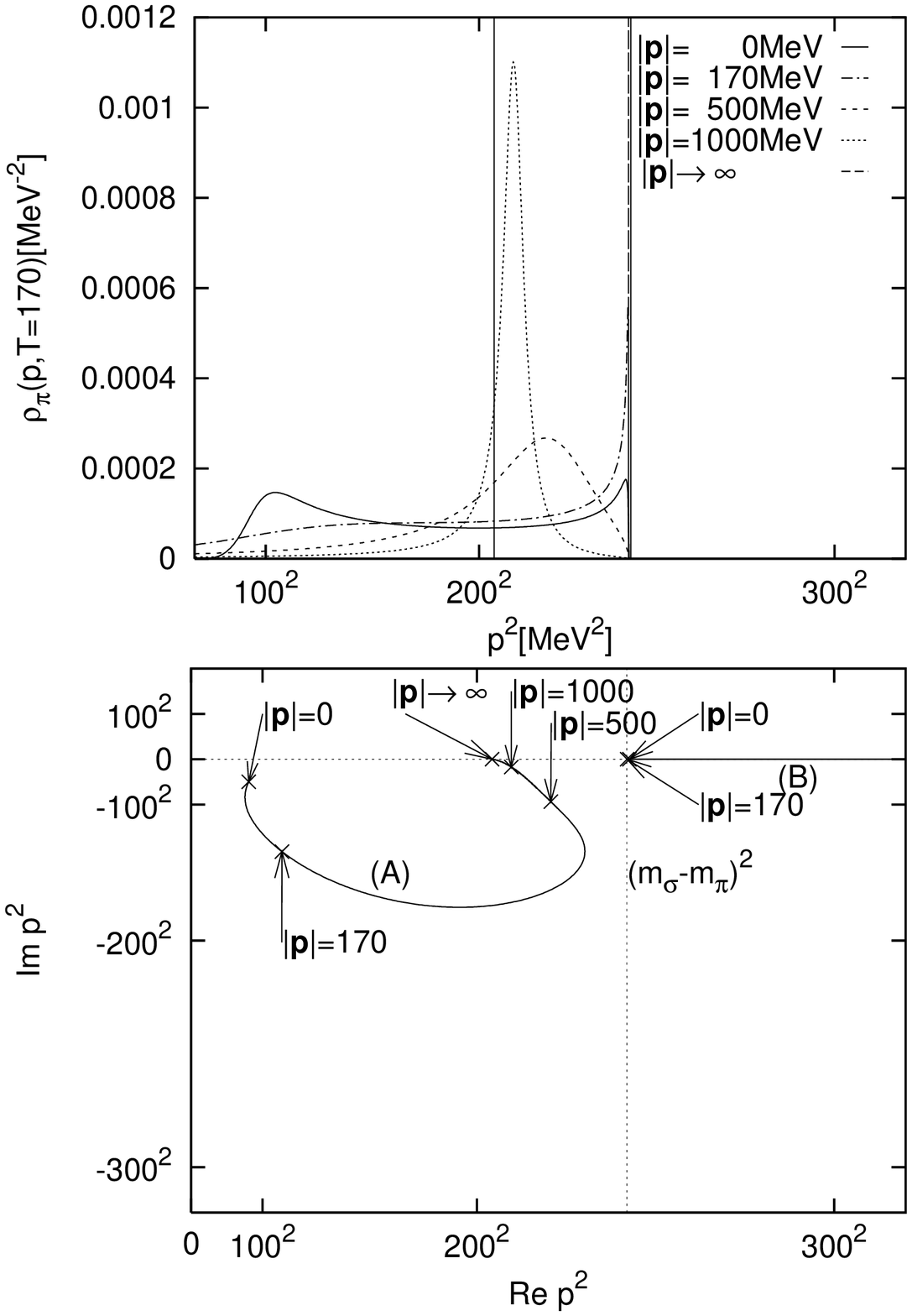} 
    \caption{Spectral function and the pole position of propagator at $T=170$MeV for $|\bm{p}|$=0, 170, 500, 1000, and $|\bm{p}|\rightarrow \infty$ in the $\pi$ channel.}
    \label{pdependenceT=170pi}
\end{figure}
\begin{figure}
    \includegraphics[width=.9\linewidth,clip]{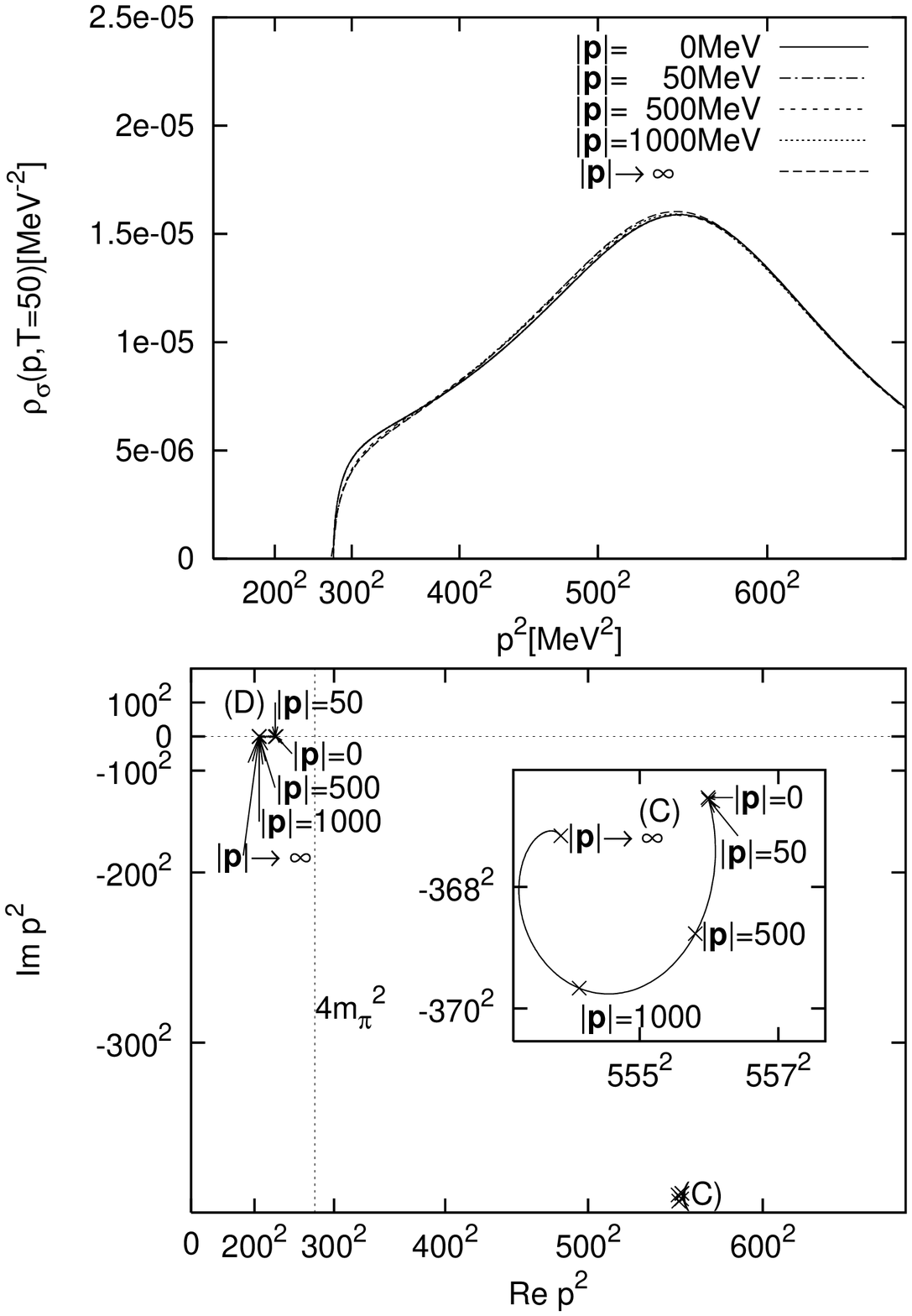}
    \caption{Spectral function and the pole position of propagator at $T=50$MeV for $|\bm{p}|$=0, 50, 500, 1000, and $|\bm{p}|\rightarrow \infty$ in the $\sigma$ channel.}
    \label{pdependenceT=50sigma}
\end{figure}

\begin{figure}
    \includegraphics[width=.9\linewidth,clip]{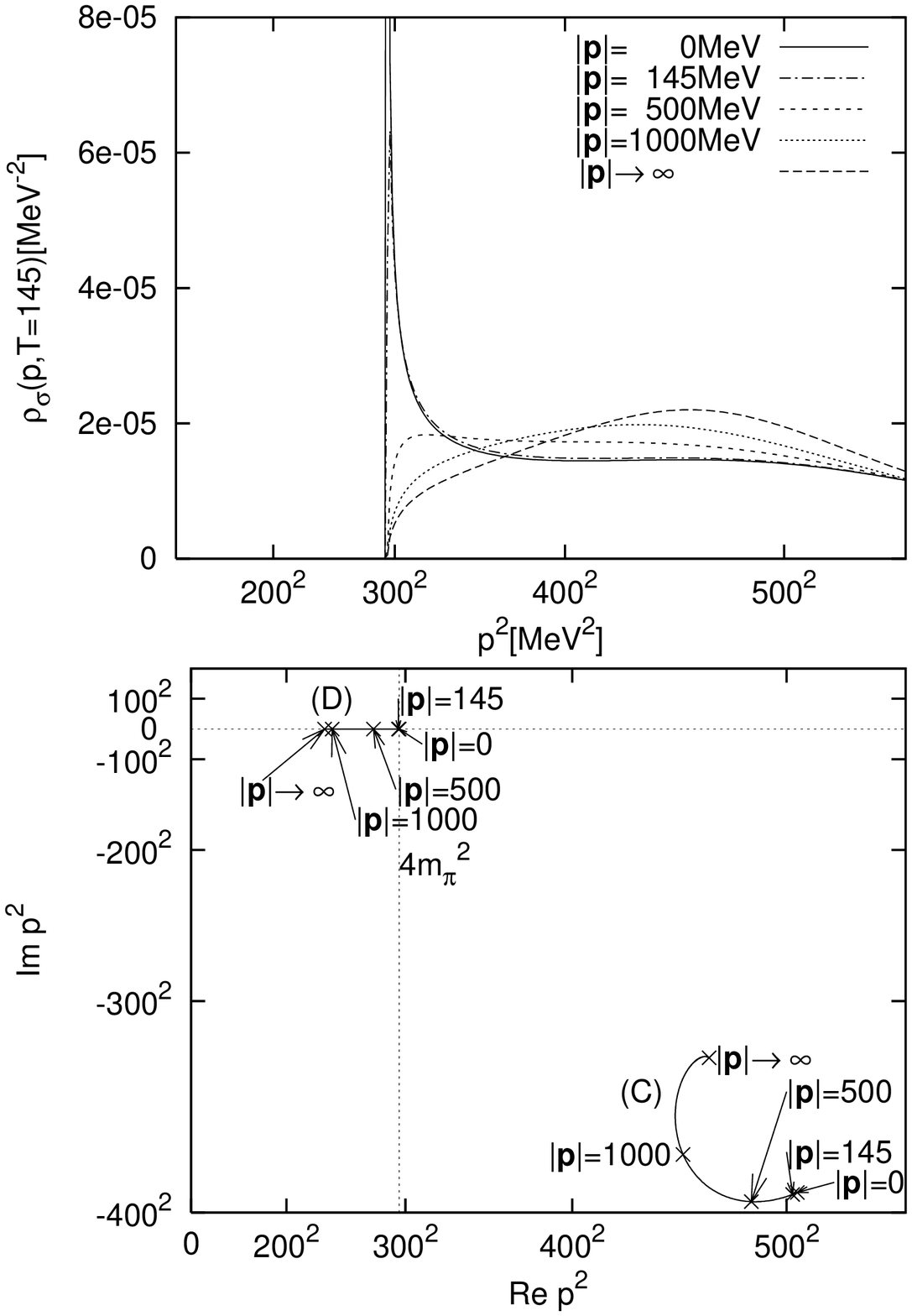}
    \caption{Spectral function and the pole position of propagator at $T=145$MeV for $|\bm{p}|$=0, 145, 500, 1000, and $|\bm{p}|\rightarrow \infty$ in the $\sigma$ channel.}
    \label{pdependenceT=145sigma}
\end{figure}

\begin{figure}
    \includegraphics[width=.9\linewidth,clip]{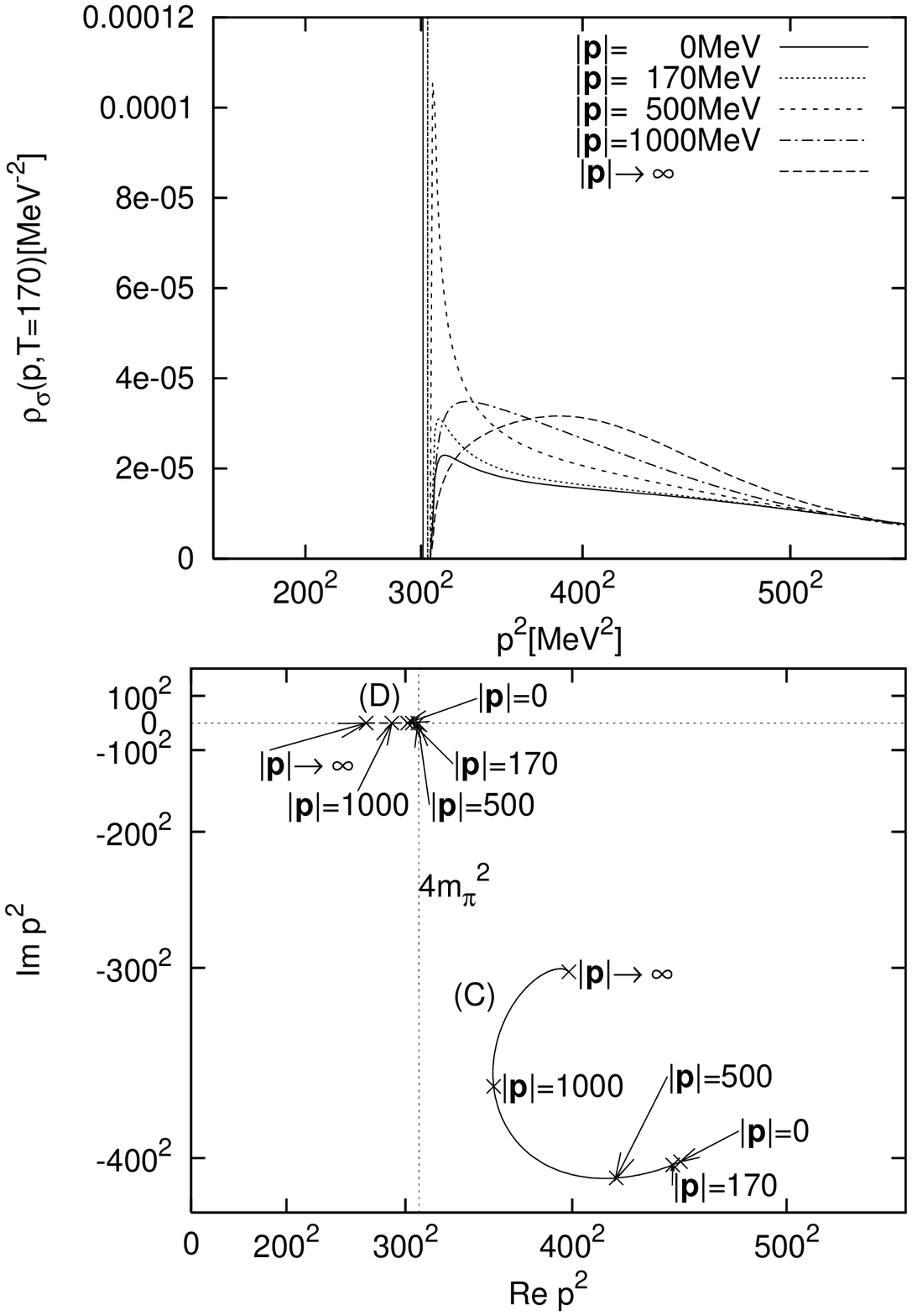}
    \caption{Spectral function and the pole position of propagator at $T=170$MeV for $|\bm{p}|$=0, 170, 500, 1000, and $|\bm{p}|\rightarrow \infty$ in the $\sigma$ channel.}
    \label{pdependenceT=170sigma}
\end{figure}
\end{document}